\documentclass[preprint,12pt]{elsarticle}
\usepackage{caption,multirow}
\usepackage{subcaption}
\usepackage{graphicx}
\usepackage{amsmath,amssymb,amsfonts}
\usepackage{algorithmic,xcolor}
\usepackage{hyperref}                                   

\journal{Engineering Applications of Artificial Intelligence}
\begin{document}
\begin{frontmatter}



\title{Elastic Net Regularization and Gabor Dictionary for Classification of Heart Sound Signals using Deep Learning}
\author{Mahmoud Fakhry\fnref{label1,label2}}
\ead{mamahmou@ing.uc3m.es}
\author{Ascensión Gallardo-Antolín\fnref{label1}}
\ead{mamahmou@ing.uc3m.es}

\affiliation[label1]{organization={Department of Signal Theory and Communications, Universidad Carlos III de Madrid},
             addressline={Avda. de la Universidad, 30},
             city={Leganés (Madrid)},
             postcode={28911},
             country={Spain}}

\affiliation[label2]{organization={Department of Electrical Engineering, Faculty of Engineering, Aswan University},
             city={Aswan},
             postcode={81542},
             country={Egypt}}
\begin{abstract}
In this article, we propose the optimization of the resolution of time-frequency atoms and the regularization of fitting models to obtain better representations of heart sound signals. This is done by evaluating the classification performance of deep learning (DL) networks in discriminating five heart valvular conditions based on a new class of time-frequency feature matrices derived from the fitting models. We inspect several combinations of resolution and regularization, and the optimal one is that provides the highest performance. To this end, a fitting model is obtained based on a heart sound signal and an overcomplete dictionary of Gabor atoms using elastic net regularization of linear models. We consider two different DL architectures, the first mainly consisting of a 1D convolutional neural network (CNN) layer and a long short-term memory (LSTM) layer, while the second is composed of 1D and 2D CNN layers followed by an LSTM layer. The networks are trained with two algorithms, namely stochastic gradient descent with momentum (SGDM) and adaptive moment (ADAM). Extensive experimentation has been conducted using a database containing heart sound signals of five heart valvular conditions. The best classification accuracy of $98.95\%$ is achieved with the second architecture when trained with ADAM and feature matrices derived from optimal models obtained with a Gabor dictionary consisting of atoms with high-time low-frequency resolution and imposing sparsity on the models.
\end{abstract}


\end{frontmatter}


\section{Introduction}\label{sec1}
An estimated 17.9 million people die each year from cardiovascular diseases (CVDs) \cite{www11a}, {with a third of these deaths occurring prematurely in people under $70$}. These diseases are a group of heart and blood vessel disorders that include coronary heart disease, cerebrovascular disease, and rheumatic heart disease. 

Several CVDs can produce changes in heart sounds. Cardiologists are used to examining heart health by hearing its sound \cite{Ranganathan2015TheAA}. This traditional examination is time consuming and requires extensive knowledge to be learned over many years. For an accurate diagnosis of CVDs, it was the beginning of thinking of automated examination of heart health. Electrocardiogram (ECG) and photoplethysmogram (PPG) signals are used for such medical examination \cite{Dutt2015DigitalPO}. The phonocardiogram (PCG) signal, the recording of sounds and murmurs made by the heart, can also be exploited for such task \cite{Sprague1954TheCV,mfakhryvmd}. The ECG and PCG signals are highly correlated and are known to contain more information than the PPG signal. {The PCG signal, however, enjoys a distinct advantage over the ECG signal as it records the acoustic properties of heart chambers \cite{Debbal2008ComputerizedHS}.} Starting from this point, the main focus of this paper is on the development of an automatic diagnosis system of CVDs through optimized modeling of PCG signals and deep learning techniques.

In one cardiac cycle of PCG signals, the predominant auscultation findings are two heartbeats and extra murmurs. Heartbeats are short transient signals produced by valve opening and closing and are called heartbeats $S_1$ and $S_2$. Murmurs last longer than heartbeats and are caused by the turbulence of blood flow. For the purpose of automatic examination of heart health, many analysis techniques have been developed to measure and quantify the features of PCG signals \cite{Abbas2009PhonocardiographySP,Fakhry2022ComparisonOW}, such as the timing of the heartbeats and their frequency components, their locations in the cardiac cycle, and the envelope shape of the murmurs. Concerning the nonstationary nature of PCG signals, they can be well analyzed and evaluated to obtain such valuable features by integrating both time and frequency representations in one representation, which is known as time-frequency analysis. This is conventionally done, for example, using the short-time Fourier transform (STFT) {or a variant of the wavelet transform (WT), mainly the continuous wavelet transform (CWT) or the discrete wavelet transform (DWT)}
\cite{Graps1995AnIT,ElAsir1996TimefrequencyAO}. In line with traditional time-frequency analysis techniques, an overcomplete analysis dictionary has recently been exploited for signal analysis \cite{771038}.

\textit{Analysis dictionaries} composed of column vectors of \textit{time-frequency atoms} play an important role in signal modeling and analysis. In their design, two competing factors warrant consideration. Any dictionary should be complete for the class of signals under inspection; however, at the same time, it is impractical to have unnecessarily high redundancy overall. Thus, it is often desirable to employ prior knowledge concerning the signal structure to build an optimal dictionary. It is obvious that an optimal dictionary is one that allows the most successful representation of a prominent PCG signal with minimal computational error and cost \cite{Koymen1987ASO,Zhang1998AnalysissynthesisOT}. Among other alternatives, the optimal Gabor dictionary has several advantages in analyzing transient heartbeats, as it ensures rapid and efficient implementation \cite{Qiu1995DiscreteGS}. \textbf{Our aim is to explore different time-frequency resolutions for atoms that may be used to build an optimal Gabor dictionary}. In particular, we propose starting from an arbitrary \textit{Gabor dictionary} and optimizing its entries for the inspected PCG signals. 

A major goal in time series modeling and learning problems is to select from the analysis dictionary a few time-frequency atoms that are suitable to efficiently represent PCG signals. In this context, {this process typically consists of determining} a linear combination of those atoms so that the measured error between the PCG signal and the combination of atoms is minimized. This linear combination of atoms is defined by \textit{a coefficient vector}, where each entry describes the contribution of the corresponding time-frequency atom in the analysis dictionary. It is widely accepted that the number of atoms in the dictionary is usually very large, possibly much larger than the number of atoms selected to represent PCG signals. 

Atom selection is tackled by examining all possible subsets of atoms; although theoretically appealing, it is computationally unfeasible. One strategy to overcome this issue makes use of \textit{greedy matching pursuit (MP)}, which approximates the minimizer of the measured error penalized with the nonconvex $l_0$ norm of the coefficient vector \cite{Mallat1993MatchingPW}. Another approach is the adoption of sparsity-based regularization schemes through \textit{the least absolute shrinkage and selection operator} (LASSO) {that use the convex $l_1$ norm instead of the $l_0$ norm} \cite{Tibshirani1996RegressionSA}. LASSO has some drawbacks in atom selection, where there are highly correlated atoms, and we need to identify all relevant ones. This motivated the use of a penalty that is a weighted sum of the $l_1$ norm and the square of the $l_2$ norm of the coefficient vector. A regularization parameter controls the amount of sparsity enforced by the $l_1$ norm and {handles multicollinearity in the dataset enforced by the squared $l_2$ norm}. The corresponding method is called \textit{elastic net regularization of linear models} \cite{Zou2005RegularizationAV}. \textbf{Our aim is to explore different values for the regularization parameter that may provide an efficient estimation of a coefficient vector associated with a defined dictionary for representing a PCG signal}. We propose examining a regularization parameter with values between zero that corresponds to the ridge regression and one that corresponds to LASSO.

{Regarding the automatic classification of CVDs,} recently deep neural networks have shown superiority {for this task} in comparison to traditional methods \cite{Gelpud2021DeepLF,Alkhodari2021ConvolutionalAR}. {blue}{Deep classifiers can include convolution neural networks (CNN) \cite{Gelpud2021DeepLF}, recurrent neural networks (RNNs) \cite{Alkhodari2021ConvolutionalAR}, or combinations of both (CNN-RNN) \cite{Alkhodari2021ConvolutionalAR,AlIssa2022ALH}}. Convolutional layers are able to extract representative information \cite{LeCun2015DeepL}, and recurrent layers, such as the long short-term memory (LSTM) layer, connect hidden nodes to create a cycle, allowing the output of some nodes to affect the input of the same nodes \cite{LeCun2015DeepL}, which is very convenient for modeling temporal sequences, as PCG signals. {In the aforementioned studies using deep classifiers, the input to the model was the STFT \cite{AlIssa2022ALH}, DWT \cite{Alkhodari2021ConvolutionalAR}, or CWT \cite{Gelpud2021DeepLF} features of the PCG signals.} On the contrary, in this work, we propose a deep neural network composed of multiple convolutional and recurrent layers where the features are derived from the coefficient vectors. However, these vectors need to be transformed to be effectively used as input to these kinds of network architectures. Furthermore, we apply a weighted logarithmic function to calculate the features \cite{mfakhryvmd}. \textbf{For this purpose, we propose reshaping each coefficient vector to form a $2$-D time-frequency matrix that represents its corresponding PCG signal in such a way that it could be used to feed a CNN-LSTM. Furthermore, our aim is to conduct an extensive evaluation of the performance of the CNN-LSTM model-based classifier with these time-frequency feature matrices as input for the diagnosis of CVDs}.

On the basis of the above discussion, the novel contribution of this work is concluded as follows.
\begin{enumerate}
    \item Analyze the modeling performance of various dictionaries, each with a specific resolution for time-frequency atoms. This is done to find an optimal dictionary that accurately represents the PCG signals under examination.
    \item Examine various settings for the regularization parameter in the elastic net regularization model. We do this to find the optimal value of the regularization parameter that can be used in conjunction with the optimal dictionary to model the PCG signals under investigation.
    \item From the above two steps, we aim to estimate coefficient vectors with high discrimination properties. We propose developing a new class of time-frequency feature matrices derived from those coefficient vectors {that will be used to build} a deep neural network classifier for the diagnosis of CVDs.
\end{enumerate}
\section{Heart sound signals}
The human heart consists of four chambers. Two of them are known as atria, and they form the upper part of the heart \cite{humanheart2007}. The lower portion of the heart is made up of the two remaining chambers, which are known as ventricles. Mitral and tricuspid valves are located, respectively, between the left atrium and left ventricle and the right atrium and right ventricle. Blood enters the atria through the aortic valve, located between the left ventricle and the aorta. Blood exits the ventricles through the pulmonary valve located between the right ventricle and the pulmonary artery. As the heart muscle contracts and relaxes, the valves open and close, allowing blood to flow into the ventricles and atria at alternate times. 

{Heart valves can experience two types of malfunction: \textit{regurgitation and stenosis}. Regurgitation occurs when the valve does not close fully, allowing blood to flow backward.} {Stenosis indicates that there is a narrowing of the valve or the valve becomes damaged or scarred, inhibiting the flow of blood out of the ventricles or atria.}

The cardiac cycle comprises two phases: systole, during which the ventricles contract to pump blood out, and diastole, during which they relax and fill with blood. {Two heartbeats, known as the first heartbeat ($S_1$) and the second heartbeat ($S_2$), are made by the heart of a healthy adult.} However, other sounds and murmurs can also be present. In general, major auscultation findings can be classified as
\begin{itemize}
    \item \textbf{Systolic heart sounds} include the first heartbeat ($S_1$) and clicks. $S_1$ occurs at the beginning of systole due to mitral closure, but may also include components of tricuspid closure. Clicks occur at systole; they are distinguished from $S_1$ by their higher pitch and briefer duration. 
    \item \textbf{Diastolic heart sounds} include $2^{nd}$, $3^{rd}$, and $4^{th}$ heartbeats ($S_2$, $S_3$, and $S_4$). $S_2$ occurs at the beginning of diastole, due to closure of the aortic and pulmonary valves. $S_3$ occurs at the early stage of diastole during filling of the diastolic ventricular and indicates serious ventricular dysfunction. $S_4$ is produced by augmented ventricular filling, caused by atrial contraction, near the end of diastole. $S_4$ is heard much more than $S_3$ and indicates a lesser degree of ventricular dysfunction. 
    \item \textbf{Systolic murmurs} may be normal or abnormal; early, mid, or late systolic. They may be divided into ejection, regurgitation, and shunt murmurs. Ejection murmurs are due to turbulent forward flow through irregular valves. Regurgitation murmurs represent the abnormal flow into heart chambers that are at a lower resistance. Shunt murmurs may originate at the site of the shunt or result from altered hemodynamics remote from the shunt.
    \item \textbf{Diastolic murmurs} are always abnormal; most are early or mid-diastolic, but can be late diastolic. Early diastolic murmurs are usually caused by aortic regurgitation or pulmonic regurgitation. Mid-diastolic murmurs are typically due to mitral or tricuspid stenosis. Rheumatic mitral stenosis may cause a late diastolic murmur.
    \item \textbf{Continuous murmurs} are always abnormal, indicating constant shunt flow throughout systole and diastole. They may be associated with signs of right and left ventricular hypertrophies. As the resistance of the pulmonary artery increases in shunt lesions, the diastolic component gradually decreases. When the pulmonary and systemic resistance equalizes, the murmur may disappear.
\end{itemize}

\subsection{Dataset}
\label{subsec:dataset}
\begin{figure*}[hbt!]
\centering
\includegraphics[scale=0.68,trim={0cm 5cm 0cm 5cm}]{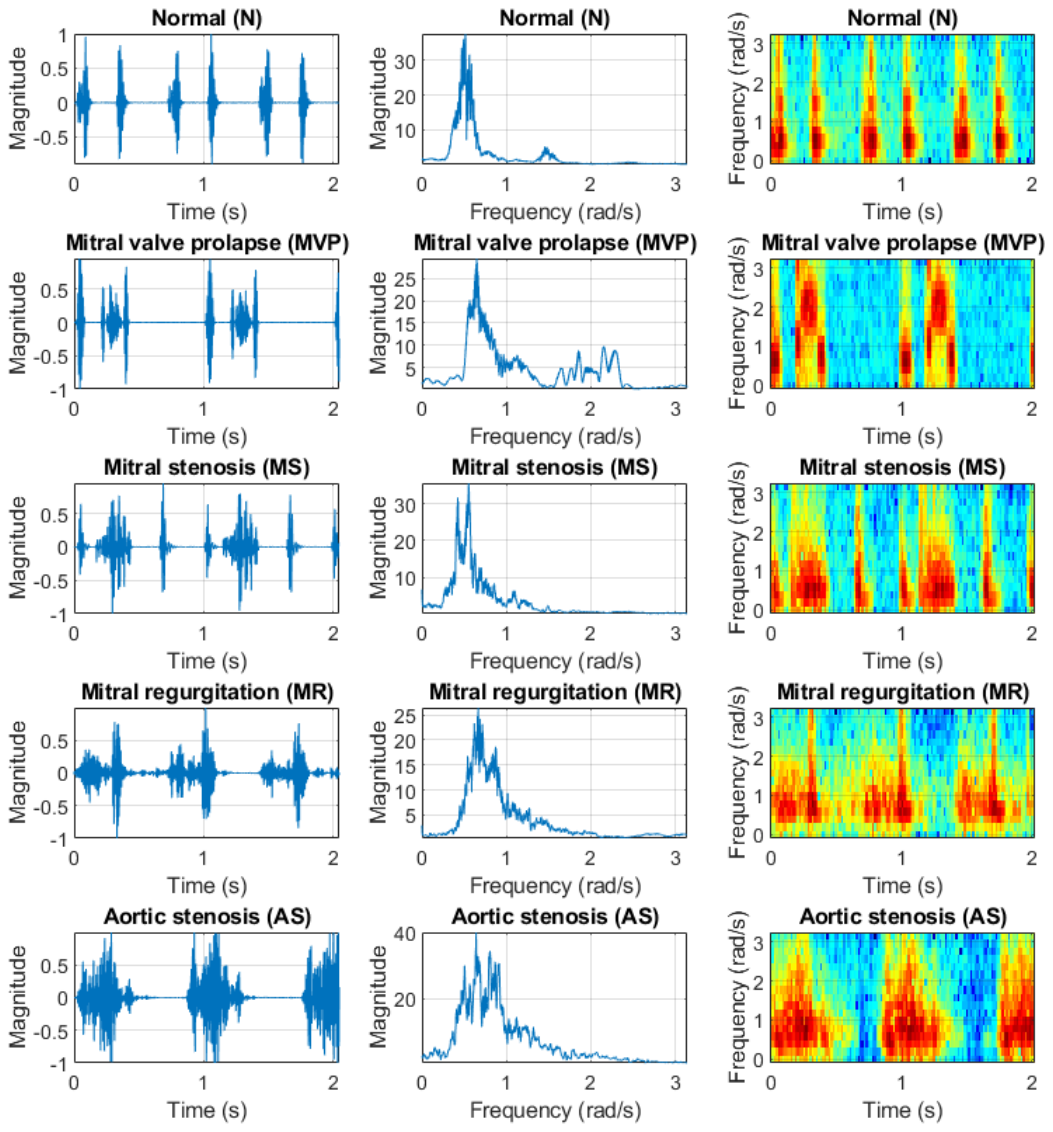}
\caption{Plots of heart sound signals and their corresponding Fourier transforms and spectrograms at a sampling frequency of $f_s=1000$ Hz, for a healthy person and four patients with CVDs. The signal length is $2$ seconds and the frequency is between $0$ and $\pi$.} 
\label{fig:cardiac02}
\end{figure*}

The dataset used in our experiments has recently been utilized by many scholars \cite{Alkhodari2021ConvolutionalAR,AlIssa2022ALH,app8122344,Jain2022AL1}. The average duration of the recordings is almost $20000$ samples or $2.5$ seconds at a sampling rate of $8000$ Hz. The recordings are from five categories, {one normal and four abnormal classes: mitral valve prolapse, mitral stenosis, mitral regurgitation, and aortic stenosis.} The total number of recordings is $1000$ ($200$ audio files/category). 
\begin{itemize}
    \item \textbf{Normal (N)} refers to the PCG signals of healthy hearts. They consist of the first heartbeat ($S_1$), the second heartbeat ($S_2$) and the silent systolic and diastolic intervals.
    \item \textbf{Mitral valve prolapse (MVP)} is the systolic prolapse of the mitral leaflet in the left atrium. Although MVP is generally not harmful, it can have side effects such as chordal rupture, endocarditis, and mitral regurgitation. If regurgitation is substantially present, signs include a mid-systolic click and a late-systolic murmur.
    \item \textbf{Mitral stenosis (MS)} occurs when the mitral valve is unable to fully open. This makes the left atrium struggle to pump blood into the left ventricle, causing blood to pool in the pulmonary circulation. Heart sounds will reveal accentuated $S_1$ early in MS and soft $S_1$ in severe MS. $S_2$ becomes more prominent as pulmonary hypertension develops and $S_3$ is almost absent in pure mitral stenosis. MS also typically produces mid and late-diastolic murmurs in the case of rheumatic MS.
    \item \textbf{Mitral regurgitation (MR)} occurs when the mitral valve does not close completely, causing blood to return to the heart rather than being pumped out. $S_1$ may be soft or even absent in MR due to leaflet sclerosis of the valve. MR also produces murmurs that begin after $S_1$ under conditions that cause leaf absence during systole, and they get louder till $S_2$. A short mid-diastolic murmur can also be heard due to the torrential mitral outflow after $S_3$.    
    \item \textbf{Aortic stenosis (AS)} happens when the aortic valve is tiny, narrow, or stiff, resulting in late closure of the aortic valve. High-pitched diamond-shaped murmurs are the classic symptoms of AS. Those systolic murmurs are best heard at the upper right border of the sternum, and their peaks appear during early systole in mild aortic coarctation.
\end{itemize}
All PCG recordings are temporal clipping or zero padding to the same length of $2^{14}$ samples at a sampling rate of $8000$ Hz. This step is followed by downsampling with a factor of $2^3$ to reduce computational cost. In this way, the length of each preprocessed PCG signal is $L=2^{11}$ and its sampling rate is $f_s=1000$ Hz. Finally, each PCG signal is standardized so that its mean is $0$ and its standard deviation is $1$. Figure \ref{fig:cardiac02} shows plots of heart sound signals and their corresponding Fourier transforms and spectrograms for a normal PCG signal of a healthy heart and PCG signals for four patients with different CVDs, in particular, mitral valve prolapse, mitral stenosis, mitral regurgitation, and aortic stenosis.

\section{Related works}
Heart sound signals are considered nonstationary signals because they exhibit marked changes with time and frequency \cite{Lee1999ComparisonBS,Xu2000NonlinearTC}. To understand the exact features of PCG signals, it is important to study their time–frequency characteristics \cite{Ergen2012TimefrequencyAO,Wang2001AnalysisOT}. It is shown that the instantaneous changes in time and frequency can be seen in the heartbeats $S_1$ and $S_2$ and the silent systolic and diastolic intervals with respect to the pathological abnormalities \cite{Guo1994ComparisonOT,Ghosh2021ASO}. 

Although time-frequency analysis based on the Fourier transform, such as the STFT, is a basic tool for signal analysis, it cannot track sudden changes in the time domain \cite{Rioul1992TimescaleED}. {Another drawback of the STFT is that once the analysis window size is chosen, the resolution is constant over the whole time-frequency plane.} If the analysis window is made short enough to capture rapid changes in the signal, it becomes impossible to resolve the frequency components of the signal, which are close in frequency within the analysis windows. On the other hand, if the window is made long enough to permit good frequency resolution, it becomes difficult to determine where in time the various frequency components act. {As a consequence, there is a resolution trade-off due to the fixed window used in the STFT computation \cite{Guo1994ComparisonOT}}. 

An alternative choice to overcome {the limitations of the STFT is to use the WT \cite{Graps1995AnIT}, that} has demonstrated the ability to analyze PCG signals more accurately than the STFT \cite{Lee1999ComparisonBS,Abbas2008MitralRP,Meziani2012AnalysisOP}, {as} is designed to give good time resolution at high frequencies and good frequency resolution at low frequencies. Several wavelet functions have been tested for the analysis and synthesis of PCG signals, including the DWT presented in \cite{Bertrand1994TimefrequencyDF,Senhadji1995ComparingWT} and the CWT {proposed} in \cite{Patidar2012ACW,Cherif2021ComparisonBA}.

The decomposition over families of localized atoms in time and frequency has been explored by researchers to find particular properties in different signal classes. Atoms of damped sinusoidal functions had the potential for time-frequency analysis of PCG signals \cite{Koymen1987ASO}, {that, although they} gave good reconstruction results for short transient heartbeats, they do not model long-duration heart murmurs \cite{Zhang1998AnalysissynthesisOT}. Alternatively, atoms of Gaussian function with a varying standard deviation and time-transition can be used to {model the} PCG signals \cite{Zhang1998AnalysissynthesisOT,Jabbari2011ModelingOH}. The advantage of this formulation is its ability to analyze signals with a longer time duration. In other works, a complete redundant dictionary of time-frequency atoms is constructed by scaling, translating, and frequency modulating a normalized window, {that is the Gaussian function, in the case of the Gabor analysis} \cite{Gabor1946TheoryOC,mfakhry}. 

The decomposition of PCG signals using an analysis dictionary of atoms like the ones aforementioned, combined with matching pursuit (MP) can provide a flexible time-frequency analysis technique \cite{Zhang1998TimefrequencyST,mfakhry}. MP is an iterative greedy algorithm aimed at finding a sparse combination of atoms that extract high-level features \cite{Mallat1993MatchingPW}. For example, in \cite{Sava1998ApplicationOT} the authors evaluated the performance of MP in detecting each cardiac cycle of a PCG signal and extracting average cardiac cycles. MP combined with a Gabor dictionary has also been used for the extraction of features from PCG signals for classification purposes, as in \cite{Jabbari2011ModelingOH,IbarraHernndez2018ABO}. These works considered traditional machine learning methods for the discrimination of valvular diseases, such as multilayer perceptron (MLP) \cite{Jabbari2011ModelingOH}, and support vector machine (SVM) \cite{IbarraHernndez2018ABO}.

More recently, deep learning techniques have been proposed for {tasks related to PCG signals.} {In \cite{Li2017ClassificationOH,Demir2019TowardsTC,Khan2022CardiNetAD,Khan2020DeepLB} heart sound signals were classified as normal or abnormal using STFT-based spectrograms and CNN models.} {In \cite{Sugiyarto2021ClassificationOH} features derived from the Morlet CWT were used as input to a CNN for heart valvular disorders discrimination.}
The classification of fundamental heartbeats $S_1$ and $S_2$ using CWT scalograms and a CNN was investigated in \cite{Meintjes2018FundamentalHS}. {In \cite{Jain2022AL1} the authors proposed a lightweight 1D-CNN architecture fed with DWT features to classify heart sound signals into five categories. Time-frequency feature matrices were extracted using DWT and CWT in \cite{Gelpud2021DeepLF} and a CNN model was built over them to categorize the signals as either normal or abnormal. In \cite{Alkhodari2021ConvolutionalAR,AlIssa2022ALH} the authors used characteristics extracted by means of the DWT and the STFT, respectively, to discriminate five heart valvular conditions. The classifiers consisted of the combination of CNN layers with RNN layers, such as long short-term memory (LSTM) in \cite{AlIssa2022ALH} and bidirectional LSTM (biLSTM) in \cite{Alkhodari2021ConvolutionalAR}.}

Unlike the few conventional works that used the Gabor dictionary in conjunction with the matching pursuit for feature extraction from PCG signals \cite{Jabbari2011ModelingOH,IbarraHernndez2018ABO}, we propose integrating the Gabor dictionary and the elastic net regularization of linear models into a single framework for the same purpose. Our method operates on the whole PCG signal as a single segment, whereas those traditional methods need an additional costly segmentation step to identify the boundaries of each cardiac cycle in the PCG signal. This {segmentation process} is done in \cite{Jabbari2011ModelingOH} with the help of the corresponding ECG signals in \cite{IbarraHernndez2018ABO} and using the logistic regression based on hidden semi-Markov models (HSMM) applied to PCG signals \cite{7234876}. {In addition, as mentioned before, in \cite{Jabbari2011ModelingOH, IbarraHernndez2018ABO} conventional machine learning techniques (in particular, MLP and SVM, respectively) are used as classifiers.} {Instead, our approach consists of building a deep learning-based model (CNN-LSTM) fed with a time-frequency matrix, that is created by reshaping a long coefficient vector estimated over the whole PCG signal by applying the aforementioned proposed feature calculation method.} To the best of our knowledge, the combination of features extracted from analysis dictionaries and CNN and/or LSTM architectures has not been previously proposed for the task of PCG signal classification.

{On the other hand, in \cite{mfakhryvmd}, we developed a classification system called VMD-CNN-LSTM which was built mainly based on the variational mode decomposition (VMD) and a 1D CNN-LSTM network. The basic idea is to decompose PCG signals into intrinsic components that are subsequently transformed into suitable features by applying a weighted logarithmic function to them. The 1D CNN-LSTM network is then fed with these features. Unlike the work presented in \cite{mfakhryvmd} which applies the weighted logarithmic function to intrinsic components, in this paper, we propose to use the same function on coefficient vectors estimated based on the Gabor dictionary and the elastic net regularization of linear models. Furthermore, instead of directly feeding the 1D CNN-LSTM network with the output of the function, we propose to reshape this output into a matrix and use it as time-frequency features to train and evaluate a 1D+2D CNN-LSTM network.}    
\section{Method}
\begin{figure*}
\centering
\includegraphics[scale=0.14,trim={0cm 0cm 0cm 2.5cm}]{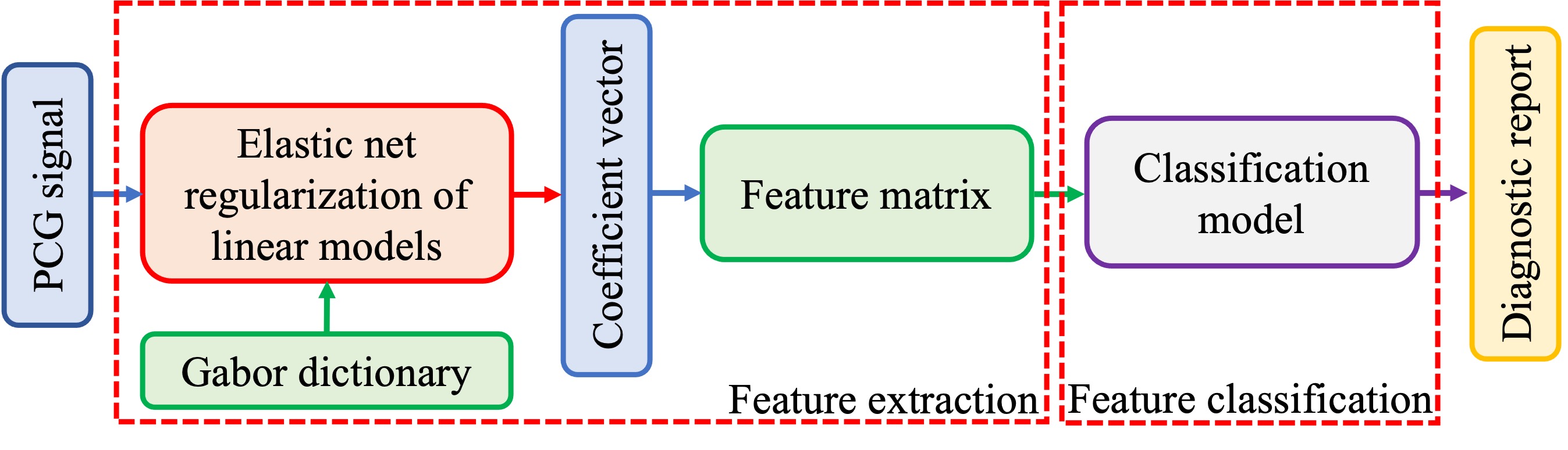}
\caption{Flowchart of the proposed method.}
\label{fig:system}
\end{figure*}

Figure \ref{fig:system} shows the block diagram of the system developed for the diagnosis of CVDs. As can be observed, it consists of two main stages: feature computation and classification. Regarding the first stage, we propose evaluating the performance of various Gabor dictionaries of time-frequency atoms in modeling PCG signals of five heart valvular conditions. This modeling task is considered as the projection of a PCG signal $\mathbf{x}$ of length $L$ onto a high-dimensional subspace or an overcomplete dictionary of normalized atoms (column vectors) $\mathbf{D}=[\mathbf{d}_1\mathbf{d}_2...\mathbf{d}_M]$ of size $L\times M$, with $L<M$. This is done using an ordinary least square term and a specifically designed sparsity-inducing regularizer. For this purpose, the best coefficient vector $\mathbf{a}$ of length $M$ is obtained by solving the following minimization problem:
\begin{equation}
\min_{\mathbf{a}} \|\mathbf{x}-\mathbf{D}\mathbf{a}\|_2 ~~ \text{subject to}   ~~ \|\mathbf{a}\|_0\leq K << M,
\label{eq:sparse}
\end{equation}
where $\|.\|_2$ is the energy norm, and the zero norm $\|.\|_0$ is the number of non-zero entries in the vector $\mathbf{a}$. This modeling allows us to acquire the most informative set of atoms. However, this depends on the construction of a good analysis dictionary and the use of an efficient regularization technique. The solution to the problem in Equation (\ref{eq:sparse}) was largely tackled using greedy matching. This is done by iteratively computing the coefficient vector $\mathbf{a}$ by projecting the residual onto the individual atoms of the dictionary. Atoms that reduce the energy of the residual the most are selected and the residual is updated. The procedure is repeated until a stopping criterion is met.

In contrast to existing works concerned with modeling PCG signals, which focus on matching pursuit combined with the Gabor dictionary, {we propose to replace the MP stage with the elastic net regularization of linear models. This way, it is possible to} control the degree of sparsity of the model and {to handle multicollinearity in the dataset by a regularization parameter}. Given a PCG signal and a dictionary with a certain resolution for its time-frequency atoms, we propose to explore different values for this regularization parameter to estimate a coefficient vector $\mathbf{a}$ considered as a model for a corresponding value. The goal of this optimization step is to estimate models with high discriminative properties that contribute to the accurate classification of PCG signals for the diagnosis of CVDs. 

As for the second stage, to successfully discriminate the five heart valvular conditions, we propose to use a deep classification model composed of CNN layers and LSTM layers. We also propose to train and test the model using time-frequency matrices derived from the estimated coefficient vectors $\mathbf{a}$ of the corresponding PCG signals. As a novel contribution, these matrices are obtained by reshaping the coefficient vectors and applying suitable mathematical functions to the entries of those vectors. 
\subsection{Gabor dictionary}
\begin{figure}
\centering
\includegraphics[scale=0.65,trim={1cm 0.5cm 0cm 2.5cm}]{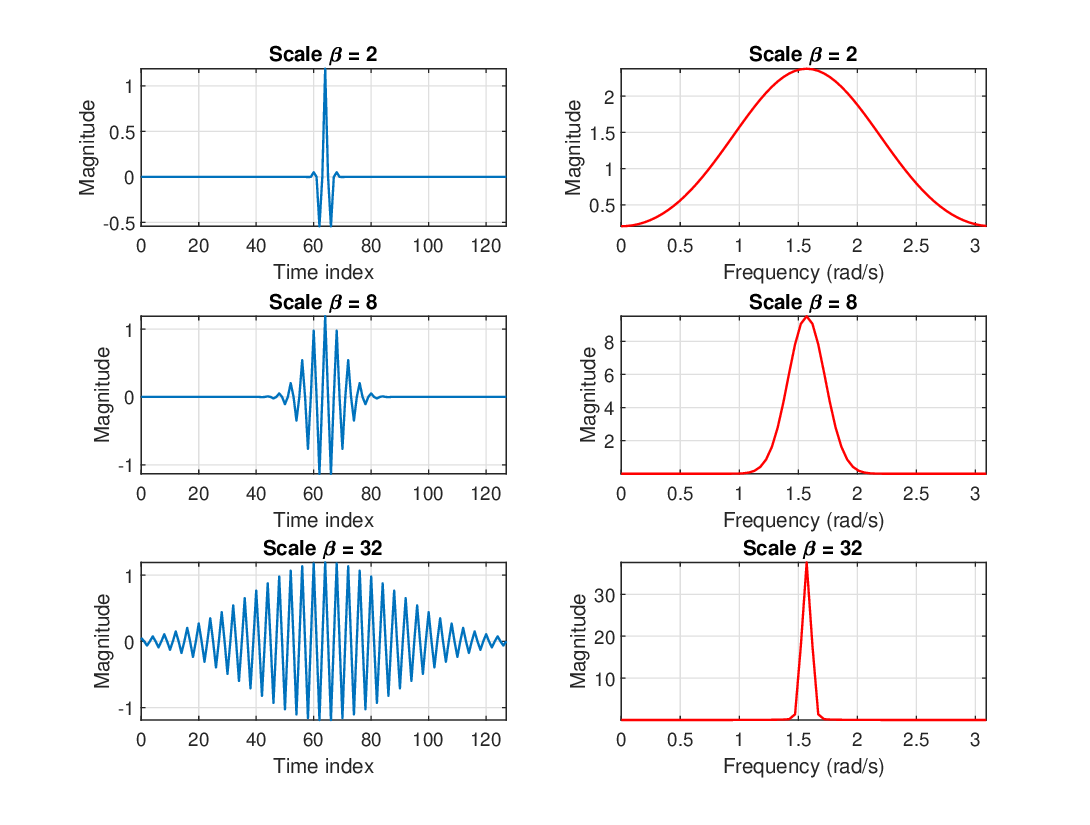}
\caption{Time and frequency representations of atoms for $3$ values for $j$ at $N=7$, $\omega=\pi/2$ and $\tau=2^6$. Increasing $\beta$ improves spectral resolution at the cost of temporal resolution.}
\label{fig:atoms}
\end{figure}

The analysis dictionary is a generalized concept for the formation of basis functions. It is common practice to use an over-complete dictionary composed of many predefined atoms, {taking into account that better localization of atoms provides better sparsity. In this work, we consider the analytic Gabor dictionary, that} is built from atoms composed of a Gaussian window function that is scaled, translated, and modulated, such as \cite{Qian1994b,Qiu1995DiscreteGS}:
 
\begin{equation}
d_{[\beta,\tau,\omega,\theta]}[n]=\frac{\gamma_{[\beta,\tau,\omega,\theta]}}{\sqrt{\beta}}g[\frac{n-\tau}{\beta}] \cos[\omega [n-\tau]+\theta],
\label{eq:atom}
\end{equation}
where $\gamma_{[\beta,\tau,\omega,\theta]}$ is a scaling constant and Gaussian function {$g[\frac{n-\tau}{\beta}]= 2^{1/4}e^{-\pi [\frac{n-\tau}{\beta}]^2}$} with $n=0,1,2,...,2^N-1$. Note that the length of the Gaussian window is $L=2^N$ and must match the length of the signal to be decomposed. To keep the overlap factor among the atoms at $50 \%$ of their duration, the scale parameter $\beta$ is first chosen with the power of $2$, then the translation parameter $\tau$ and the frequency $\omega$ are obtained as: \cite{Qiu1995DiscreteGS}
\begin{equation}
[\beta, \tau,\omega, \theta]=[2^{j},2^{j-1} p,2^{-j-1} \pi c, 0],
\end{equation}
where $0<j<N$, $p = 0:1: 2^{N-j+1}-1$, and $c = 0:1: 2^{j+1}-1$. For simplicity, the phase shift parameter $\theta$ will be assumed as equal to $0$. Consequently, the total number of atoms is given by $M= 2^{j+1} \times 2^{N-j+1} = 2^{N+2}$. The $j$th Gabor dictionary $\mathbf{D}_j$ of size $2^N \times 2^{N+2}$ is built for a specific value of $j$ by arranging the atoms side by side in matrix form. As mentioned in Subsection \ref{subsec:dataset}, all signals have the same length of $L=2^N=2^{11}$, and for that, the value of $N$ is set to $11$. Table \ref{tab:var} reports the variables of the $j$th dictionary for all possible values of $j$ at this value of $N$. 
\begin{table}
\begin{center}
\caption{Variables of the $j$th Gabor dictionary $\mathbf{D}_j$ for all values of $j$ at $N=11$ and $\theta=0$. {A discrete interval of $\omega$ or $\tau$ defined as $z_0:\delta z:z_{\infty}$ begins at $z_0$, ends at $z_{\infty}$ and is discretized with a step of $\delta z$}.}
\begin{tabular}{@{}llll@{}}
\hline
$j$ & $\beta$ & $\omega$& $\tau$\\
\hline
$1$&$2^1$&$0:\frac{\pi}{4}:\frac{3\pi}{4}$&$0:2^0:(2^{11}-2^0)$\\
$2$&$2^2$&$0:\frac{\pi}{8}:\frac{7\pi}{8}$&$0:2^1:(2^{11}-2^1)$\\
$3$&$2^3$&$0:\frac{\pi}{16}:\frac{15\pi}{16}$&$0:2^2:(2^{11}-2^2)$\\
$4$&$2^4$&$0:\frac{\pi}{32}:\frac{31\pi}{32}$&$0:2^3:(2^{11}-2^3)$\\
$5$&$2^5$&$0:\frac{\pi}{64}:\frac{63\pi}{64}$&$0:2^4:(2^{11}-2^4)$\\
$6$&$2^6$&$0:\frac{\pi}{128}:\frac{127\pi}{128}$&$0:2^5:(2^{11}-2^5)$\\
$7$&$2^7$&$0:\frac{\pi}{256}:\frac{255\pi}{256}$&$0:2^6:(2^{11}-2^6)$\\
$8$&$2^8$&$0:\frac{\pi}{512}:\frac{511\pi}{512}$&$0:2^7:(2^{11}-2^7)$\\
$9$&$2^9$&$0:\frac{\pi}{1024}:\frac{1023\pi}{1024}$&$0:2^8:(2^{11}-2^8)$\\
$10$&$2^{10}$&$0:\frac{\pi}{2048}:\frac{2047\pi}{2048}$&$0:2^9:(2^{11}-2^9)$\\
\hline
\end{tabular}
\label{tab:var}
\end{center}
\end{table}
There is a tradeoff between the temporal resolution and the spectral resolution of the dictionary, which is governed by the factor $j$; see Figure \ref{fig:atoms}. The Gaussian function {$g[\frac{n-\tau}{\beta}]$} slowly decays around translation $\tau$ as the value of $j$ increases. This increases the time duration of the atom and degrades its temporal resolution. On the other hand, the Fourier transform of the Gaussian time function is also a Gaussian function. Consequently, the bandwidth of the atom around frequency $\omega$ slowly decreases as the duration of the Gaussian function increases. This leads to atoms with better spectral resolution than atoms with shrunken Gaussian time functions. 
\subsection{Elastic net regularization of linear models}
Elastic net regularization of linear models is a broader version of ridge regression and LASSO methods. Ridge regression was developed to address the instability of estimates of the vector coefficients of linear models. Atom selection based on LASSO forces coefficients associated with some explanatory atoms to be zero and yields accurate estimates of the coefficients of the remaining atoms. For a regularization parameter $\alpha$ strictly between $0$ and $1$, and nonnegative $\lambda$ ($\lambda \ge 0$), elastic net solves the following problem \cite{Zou2005RegularizationAV}:
\begin{equation}
    \min_{\mathbf{a}_{j,\alpha}} \Big(\frac{1}{2N} \|\mathbf{x}-\mathbf{D}_j\mathbf{a}_{j,\alpha}\|_2^2 +\lambda \big(\frac{1-\alpha}{2}\|\mathbf{a}_{j,\alpha}\|_2^2+\alpha \|\mathbf{a}_{j,\alpha}\|_1)\Big).
\label{eq:elastic_net}     
\end{equation}

{In Equation (\ref{eq:elastic_net}), the extreme values $\alpha = 0$ and $\alpha = 1$ correspond, respectively, to the ridge regression and LASSO. However, when considering intermediate values of $\alpha$ (for example, $\alpha = 0.5$ gives the same weight to the ridge regression and LASSO), the model benefits both from the ability of ridge regression to handle multicollinearity in the dataset and from the feature selection capability of LASSO. A range for $\alpha$ is initially set to estimate the coefficient vector in such a way that the best model that could produce the lowest error as well as the smallest number of nonzero entries is selected. Elastic net regularization of linear models uses a geometric sequence of values for $\lambda$, which controls the shrinkage of the features. The optimal value $\lambda$ is determined based on the minimum mean squared error.} 

\begin{figure*}[hbt!]
  \centering
    \includegraphics[scale=0.90]{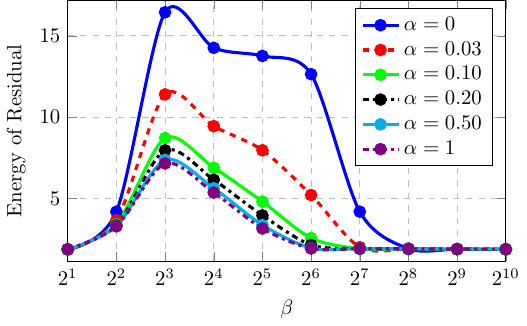}
    \caption{Approximation error measured as the average energy of all residuals $\|\mathbf{x}-\mathbf{D}_j\mathbf{a}_{j,\alpha}\|_2^2$, calculated for all the recordings of the dataset.}
  \label{fig:error1}
\end{figure*}

\begin{figure*}[hbt!]
  \centering
    \includegraphics[scale=0.90]{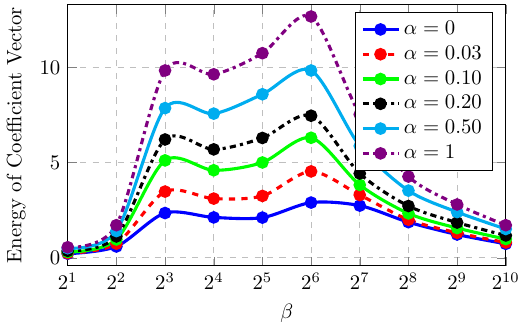}
    \caption{Average energy $\|\mathbf{a}_{j,\alpha}\|_2^2$, calculated for all the recordings of the dataset.}
      \label{fig:energy}
\end{figure*}

\begin{figure}
\centering
\includegraphics[scale=0.70,trim={0.8cm 1.5cm 0cm 2cm}]{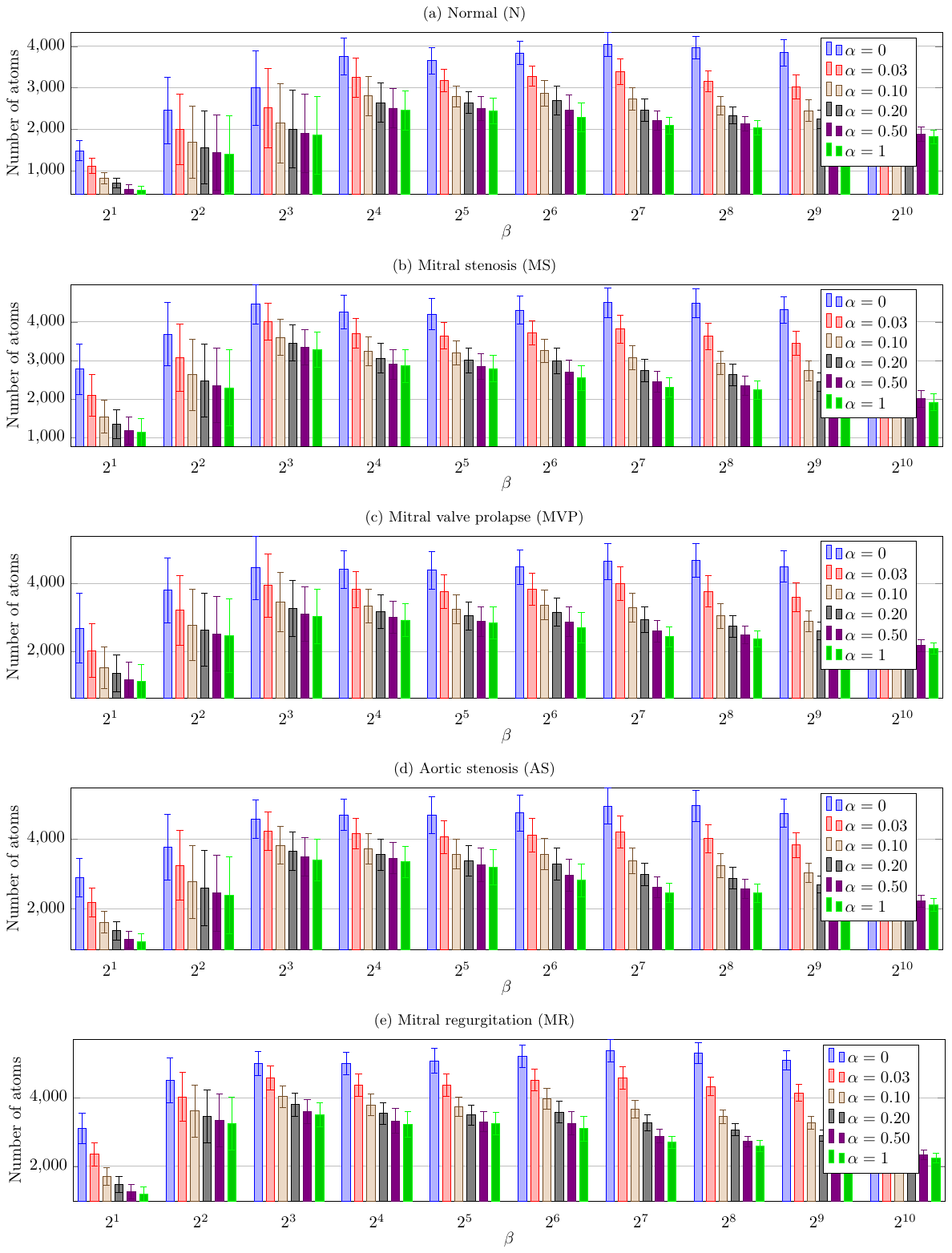}
\caption{Average number of atoms corresponding to the number of non-zero entries in $\mathbf{a}_{j,\alpha}$. These numbers and their standard deviations are computed using $200$ PCG signals for each heart condition. Signals are truncated to the length of $2^{14}$ samples, downsampled by a factor of $8$, and then approximated with a Gabor dictionary of size $2^{11} \times 2^{13}$.}
\label{fig:noATOMS}
\end{figure}

{The problem in Equation (\ref{eq:elastic_net}) is optimized using the alternating direction method of multipliers (ADMM) \cite{Boyd2011DistributedOA}. The iterative update of the vector $\mathbf{a}_{j,\alpha}$ terminates when the relative change in its size is below a threshold. At this convergence point, we can measure some quantities of interest that can help in identifying the optimal dictionary and regularization parameter from the sense of sparsity, such as the energy of approximation error, the energy of the coefficient vector, and its average number of non-zero elements, that are displayed in, respectively, Figures \ref{fig:error1},  \ref{fig:energy} and \ref{fig:noATOMS}.} 

Figures \ref{fig:error1} and \ref{fig:energy} show the average approximation error to estimate the fitting models $\mathbf{a}_{j,\alpha}$ as well as their average energy, respectively; they are plotted as functions of the scale parameter $\beta$ and the regularization parameter $\alpha$. This is done using $1000$ recordings for the five heart conditions. The error is calculated as the energy of the residual obtained as $\|\mathbf{x}-\mathbf{D}_j\mathbf{a}_{j,\alpha}\|_2^2$. As observed in Figure \ref{fig:error1}, the error is very small for dictionaries of atoms with very high-time low-frequency resolution ($\beta=2^1$) and for dictionaries of atoms with low-time high-frequency resolutions ($\beta=2^8$, $2^9$ and $2^{10}$) for all values of $\alpha$. From $\beta=2^2$, the error gradually increases as $\beta$ increases and records a high peak at $\beta=2^3$. After that peak, the error decreases as $\beta$ increases until it reaches its minimum values for $\beta>2^7$. When $\beta$ is between $2^3$ and $2^7$, LASSO ($\alpha=1$) provides a small error compared to the ridge regression ($\alpha=0$), and the error gradually decreases as $\alpha$ increases. The average energy of the vector $\mathbf{a}_{j,\alpha}$ follows a trend similar to the approximation error in terms of $\beta$ and a reversed relation in terms of $\alpha$, as observed in Figure \ref{fig:energy}. 

Figure \ref{fig:noATOMS} shows the average number of atoms corresponding to nonzero entries of the models $\mathbf{a}_{j,\alpha}$. Average numbers and standard deviations are reported in terms of {the $\beta$ and $\alpha$ parameters.} We can observe that dictionaries of atoms with very high-time low-frequency resolution ($\beta=2^1$) provide high sparse coefficient vectors, $\mathbf{a}_{j,\alpha}$ because a few numbers of atoms are needed to approximate the PCG signals. The signals of healthy hearts are approximated by the fewest number of atoms compared to those of valvular heart conditions, and this number decreases for all signals as the value of $\alpha$ increases. Dictionaries of atoms with low-time high-frequency resolutions provide high sparse approximations compared to dictionaries of atoms with moderate resolutions ($\beta=2^3$, $2^4$, $2^5$, $2^6$ and $2^7$) and less sparse approximations compared to dictionaries of atoms with very high-time low-frequency resolutions for all values of $\alpha$.

From the above discussion, we conclude that dictionaries of atoms with very high-time low-frequency resolutions (small $\beta$) provide the best approximation for all signals in the sense of approximation error, the energy of coefficient vectors, and their sparsity, regardless of the value of $\alpha$. Furthermore, dictionaries of atoms with low-time and high-frequency resolutions ($\beta=2^8$, $2^9$, and $2^{10}$) perform better than dictionaries of atoms with moderate time-frequency resolutions. However, the approximation performance depends on the value of $\alpha$, where large values of $\alpha$ outperform small ones in the sense of approximation error, the energy of coefficient vectors, and sparsity.

\subsection{Feature computation and representation}
\label{subsec:feature_computation_and_representation}
The computation and representation of features often govern the performance of classification models. Using the raw coefficient vectors $\mathbf{a}_{j,\alpha}$ is one method for calculating {these features.} Another possibility is to apply some kind of transformation to the coefficient vector $\mathbf{a}_{j,\alpha}$, such as calculating the magnitude squared or using the weighted logarithmic function, as shown in Subsection \ref{subsubsec:feature_computation} \cite{mfakhryvmd}. On the other hand, the dimensionality of the feature vector can be modified so that the classification model learns neighboring in higher dimensions. In this work, we propose to accomplish it by reshaping the 1D feature vector to form a 2D matrix, as described in Subsection \ref{subsubsec:feature_representation}.

\subsubsection{Feature computation}
\label{subsubsec:feature_computation}
{First of all, the coefficient vector $\mathbf{a}_{j,\alpha}$ is standardized to zero mean and unit variance and then is normalized to take values between $-1$ and $1$ (note that $\mathbf{a}_{j,\alpha}$ is a one-dimensional array of length $2^{N+2}$).} {The next step consists of applying a certain transformation to the coefficient vectors $\mathbf{a}_{j,\alpha}$ in order to obtain a new feature vector $\mathbf{b}_{j,\alpha}$}. As stated above, the neural network can be trained and tested using the raw coefficient vector or using their magnitude squared. In this paper, we use another option, in which the vector $\mathbf{b}_{j,\alpha}$ is defined in terms of the vector $\mathbf{a}_{j,\alpha}$ using the following weighted logarithmic function  \cite{mfakhryvmd}:
\begin{equation}
b^{m}_{j,\alpha} = -
\lvert{a}^{m}_{j,\alpha}\rvert \log
\lvert{a}^{m}_{j,\alpha}\rvert.
\label{eq:xk}
\end{equation}

\begin{figure*}[h]
\centering
\includegraphics[scale=1,trim={0cm 0cm 0cm 1cm}]{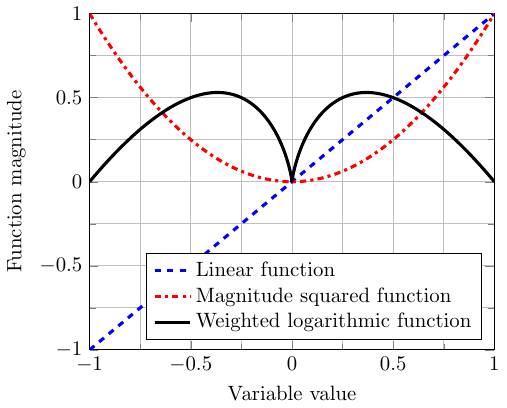}
\caption{Three different transformation functions for the calculation of features.}
\label{fig:entropy}
\end{figure*}

{Figure \ref{fig:entropy} shows plots for the linear, magnitude squared, and weighted logarithmic function in terms of a variable that takes values between $-1$ and $1$.} In \cite{mfakhryvmd} it is demonstrated that better classification performance is obtained in a CNN-LSTM network when using features calculated with the weighted logarithmic function compared to the other methods for feature computation. 

In classification models built based on CNN, the convolutional layer calculates the patterns of its input features using a set of filters or kernels $\mathbf{h}_i$, of length $Q$ with $Q<<M$ and $i=1,..., I$, where $M$ is the length of the input features and $I$ is the number of kernels considered. The convolution operation between the vector $\mathbf{b}_{j,\alpha}$ and the $i$th filter $\mathbf{h}_i$ to obtain the output $\mathbf{y}_i$ is defined as:
\begin{equation}
\mathbf{y}_i = \mathbf{h}_i * \mathbf{b}_{j,\alpha}.
\label{eq:yi}
\end{equation}

{In order to gain insight into the meaning of Equation (\ref{eq:yi}) from a statistical perspective, it can be observed that the} absolute value of the vector $\mathbf{a}_{j,\alpha}$ can be {interpreted} as a probability distribution function (PDF). Consequently, the patterns $y^n_i$ can be considered as the weighted entropy of a random variable with a probability function $\mathbf{a}_{j,\alpha}$ \cite{Kelbert2017WeightedEA}. This weighted entropy is calculated for probabilities obtained on the basis of temporal neighboring as follows:
\begin{equation}
y^n_i = -\sum^{\infty}_{v=-\infty} {h}^{n-v}_i \lvert{a}^{v}_{j,\alpha}\rvert \log \lvert{a}^{v}_{j,\alpha}\rvert.
\end{equation}
\subsubsection{Feature representation}
\label{subsubsec:feature_representation}
\begin{figure}[h]
\centering
\includegraphics[scale=0.60,trim={2.8cm 1.5cm 0cm 2.5cm}]{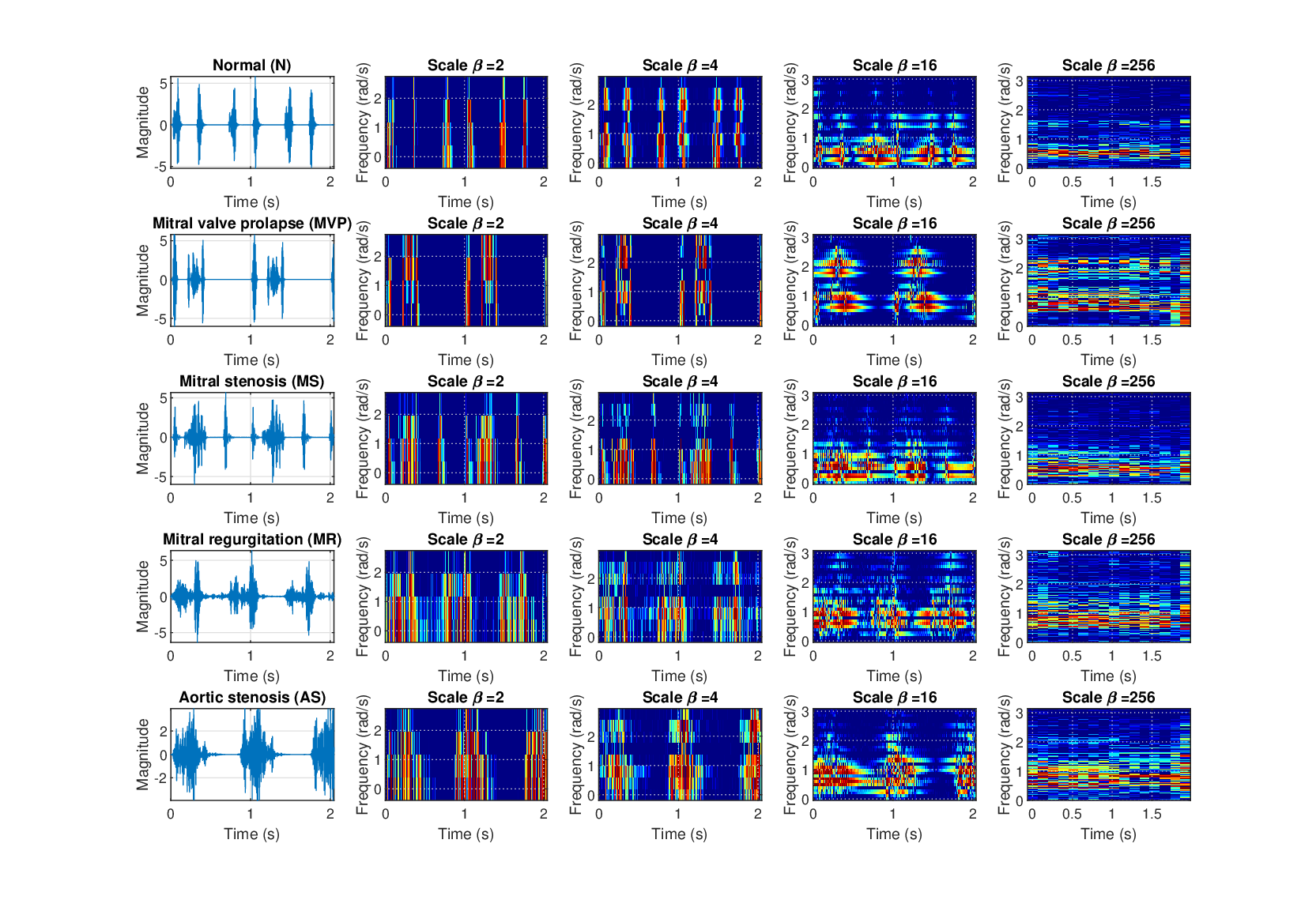}
\caption{Time-frequency feature matrices for PCG signals of five conditions. Each matrix is obtained by reshaping the weighted logarithmic values applied to the coefficient vector of the corresponding signal. Vectors are calculated using ridge regression ($\alpha = 0$) at different time-frequency resolutions for the atoms.}
\label{fig:feat1}
\end{figure}

The feature vector $\mathbf{b}_{j,\alpha}$ is of length $M=2^{N+2}$. This makes it difficult to employ such lengthy vectors as input features to classification networks. Therefore, we propose rearranging the entries of the one-dimensional vector to form a two-dimensional matrix. This is done not only to avoid dealing with long vectors but also to take advantage of modeling two-dimensional imagery data for CNN-based classification tasks. The vector of features $\mathbf{b}_{j,\alpha}$ of length $2^{N+2}$ is reshaped to build a time-frequency feature matrix $\mathbf{B}_{j,\alpha}$ of size $2^{j+1} \times 2^{N-j+1}$. Overall the translation steps $\tau$, the coefficients of the vector associated with atoms in the dictionary $\mathbf{D}_j$, having the same frequency $\omega$ are grouped together to form a row in the matrix $\mathbf{B}_{j,\alpha}$. As a result, we obtain a feature matrix with a number of rows to be $2^{j+1}$ and a number of columns to be $2^{N-j+1}$. Figure \ref{fig:feat1} shows time-frequency feature matrices for $\alpha=0$. The figure involves feature matrices of PCG signals for the five heart conditions obtained using different scaling factors $\beta$.
\subsection{CNN-LSTM classification network}
\begin{figure}%
\centering
\includegraphics[scale=0.2,trim={0cm 0cm 0cm 2cm}]{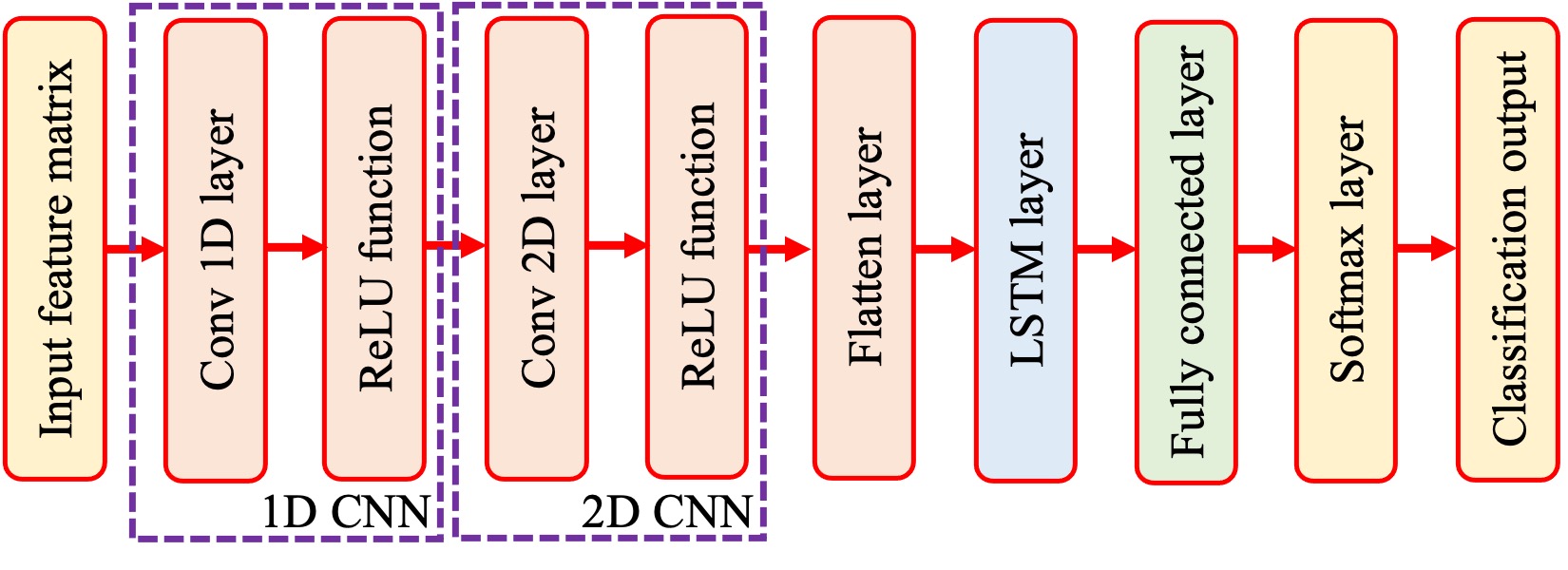}
\caption{CNN-LSTM network consisting of 1D and 2D CNN, and LSTM layers.}
\label{fig:cnnmodel}
\end{figure}

{Figure \ref{fig:cnnmodel} shows a block diagram of the architecture of the proposed CNN-LSTM model. It consists of a 1D CNN network, a 2D CNN network, and an LSTM layer, ending with dense fully connected and softmax layers used for classification purposes. It is confirmed in \cite{AlIssa2022ALH} that CNN combined with LSTM provides better results compared to CNN or LSTM alone.} CNN networks excel in extracting hierarchical structures from input data using convolution operation {with a set of filters (kernels) with specific sizes} and smoothing rectified units (ReLU) {\cite{LeCun2015DeepL}. These filters are of one dimension for a 1D CNN or two dimensions for a 2D CNN.} The 1D CNN layer deals with imagery input of two dimensions as multiple sequences of one dimension and extracts simple structures, whereas the 2D CNN layer extracts complex structures of two-dimensional imagery from the output of the 1D CNN layer. The LSTM layer excels at modeling the long-term dependencies of its input due to the presence of internal gates {\cite{Gers2003}}. The LSTM layer receives the extracted structures flattened in a flattening layer from the 2D CNN network.
\subsubsection{Network training}
The CNN-LSTM network is trained with backpropagation using an appropriate training optimization algorithm. {In particular, in this work we have performed experiments with two optimization algorithms: stochastic gradient descent with momentum (SGDM) and adaptive moment optimization (ADAM).}  SGDM relies on the gradient of the current and previous steps to identify the best course of action. This enables the loss function to get closer to its minima faster. Accordingly, momentum increases speed and accelerates convergence. ADAM modifies the learning rate using gradient decay. It takes advantage of momentum by using the moving average of the gradient rather than the gradient itself, as in the SGDM.
\subsubsection{Network implementation}
Table \ref{tab:table0000} lists the values of the parameters of the CNN-LSTM classification network. The filter and stride sizes of the 1D and 2D convolutional layers are optimized to fit the size of the matrix $\mathbf{B}_{j,\alpha}$. {As the resolution of the time-frequency feature matrices $\mathbf{B}_{j,\alpha}$ changes their size also changes. For small $j$ the matrices are wide and short, and for large $j$ the matrices are narrow and tall. Fixing the filter size will satisfy a certain resolution, but not others. Consequently, we propose using a filter of variable size that depends on $j$ so that the filter size satisfies all resolution and matrix sizes}. The table also contains the values of the training parameters of the optimization algorithms SGDM and ADAM, such as learning rate and batch size. Furthermore, we manually set two other factors, which are the momentum in SGDM and the gradient decay in ADAM. 
\begin{table*}
\centering
\caption{Implementation parameters of the CNN-LSTM classification model}
\label{tab:table0000}
\begin{tabular}{@{}ll@{}}
\hline
Parameters & Value \\
\hline
No. classes              & $5$ \\
Truncated signal length (samples)&$2^{14}$ \\
Downsampling factor      & $2^3$ \\
Downsampled signal length ($L=2^N$) & $2^{11}$\\
$j$& $1, 2, ..., 10$\\
Size of input feature matrices ($\mathbf{B}_{j,\alpha}$) & $2^{j+1} \times 2^{N-j+1}$\\
\hline
Filter size in 1D CNN    & $1 \times 2^{7-j}$, for $j \leq 5$\\
Filter size in 1D CNN    & $2^{j-4} \times 1$, for $j > 5$\\
Stride size in 1D CNN         & $1 \times 2^{6-j}$, for $j \leq 5$\\
Stride size in 1D CNN         & $2^{j-5} \times 1$, for $j > 5$\\
No. filters in 1D CNN    & $2^{6}$ \\
\hline
Filter size in 2D CNN    & $\lceil{\log_2(j)}\rceil+1 \times 2^{6-j}$, for $j \leq 5$\\
Filter size in 2D CNN    & $2^{j-5} \times 7-\lceil{\log_2(j)}\rceil$, for $j > 5$\\
Stride size in 2D CNN         & $\lceil{\log_2(j)}\rceil \times 2^{5-j}$, for $j \leq 5$\\
Stride size in 2D CNN         & $2^{j-6} \times 6-\lceil{\log_2(j)}\rceil$, for $j > 5$\\
No. filters in 2D CNN    & $2^{5}$\\
\hline
No. neurons in LSTM      & $2^{6}$\\
\hline
Learning rate in SGDM    & $0.1$\\
Momentum in SGDM         & $0.50$\\
\hline
Learning rate in ADAM    & $0.001$\\
Gradient decay in ADAM   & $0.90$\\
\hline
Batch size               & $150$\\
Max epochs               & $100$\\
\hline
\end{tabular}
\end{table*}
\subsection{Experiments}
First of all, the feature matrices of all PCG signals were extracted following the procedure explained in Subsection \ref{subsec:feature_computation_and_representation}. The experiments were performed by splitting the $1000$ extracted feature matrices into $675$ matrices to train the classification model and $325$ to test it. \textcolor{blue}{Each one of the proposed models was trained and tested $100$ times separately, using random selection for the training and testing matrices. Note that in all the experiments the data is split into disjoint training and test sets.} The final classification performance is reported by averaging the results of the hundred subexperiments.
\subsection{Performance metrics}
{In order to assess the performance of the proposed method, we have used the confusion matrix, whose} four metrics are as follows: True positive (TP) is the number of cases in which a given valvular heart disease is correctly identified. False negative (FN) is the number of cases in which the model incorrectly classifies a particular valvular heart disease as belonging to other diseases. False positive (FP) is the number of cases in which the model incorrectly classifies other diseases as belonging to a particular valvular heart disease. True negative (TN) is the number of cases in which other conditions were correctly identified. Based on these four metrics, additional metrics can be calculated to provide further information about the behavior of the model:
\begin{itemize}
\item \textbf{Precision} is the ratio between the number of correctly classified positive samples and the total number of samples classified as positive. 
\begin{equation}
\textbf{Precision} = \frac{TP}{TP+FP} \%.
\end{equation}
\item \textbf{Recall} is the ratio between the number of positive samples correctly classified as positive to the total number of positive samples. 
\begin{equation}
\textbf{Recall} = \frac{TP}{TP+FN}\%.
\end{equation}
\item \textbf{Accuracy} is calculated as the ratio between the number of correct predictions and the total number of predictions.
\begin{equation}
\textbf{Accuracy} = \frac{TP+TN}{TP+TN+FP+FN}\%.
\end{equation}
\item \textbf{Specificity} is the ratio between the number of negative samples correctly classified as negative to the total number of negative samples. 
\begin{equation}
\textbf{Specificity} = \frac{TN}{TN+FP}\%.
\end{equation}
\item \textbf{F1 score} is a harmonic mean of recall and precision. This metric was introduced to combine both metrics into a single one:
\begin{equation}
\textbf{F1 score} = \frac{2 TP}{2 TP+FP+FN}\%.
\end{equation}
\end{itemize}
\subsection{Classification results}
\begin{figure*}[h]
   \centering
   \begin{subfigure}[b]{0.45\textwidth}
   \centering   
   \includegraphics[scale=0.68]{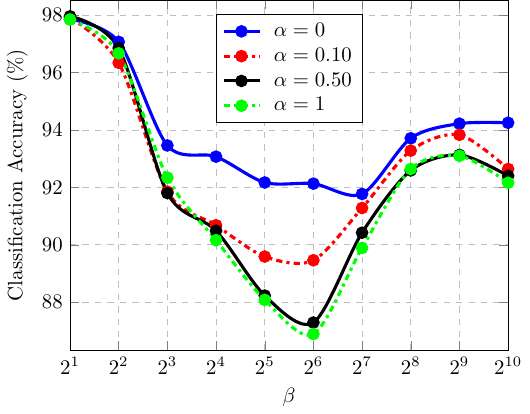}
   \subcaption{1D convolutional layer and an LSTM layer trained with ADAM.}
   \end{subfigure}
   \hfill
   \begin{subfigure}[b]{0.45\textwidth}
   \centering
   \includegraphics[scale=0.68]{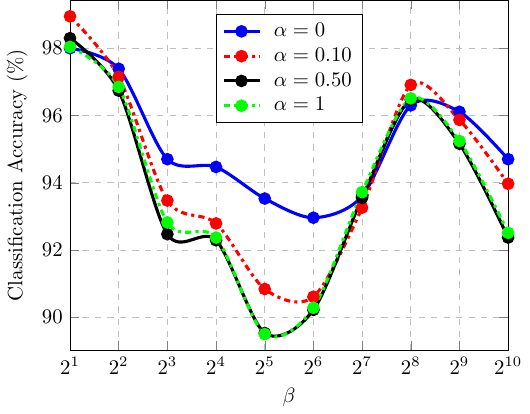}
   \subcaption{1D + 2D convolutional layers and an LSTM layer trained with ADAM.}
   \end{subfigure}
   \vfill
   \begin{subfigure}[b]{0.45\textwidth}
   \centering
   \includegraphics[scale=0.68]{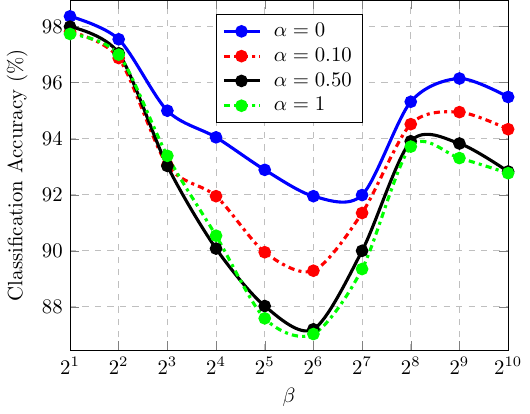}
   \caption{1D convolutional layer and an LSTM layer trained with SGDM.}
   \end{subfigure}
   \hfill
   \begin{subfigure}[b]{0.45\textwidth}
   \centering
   \includegraphics[scale=0.68]{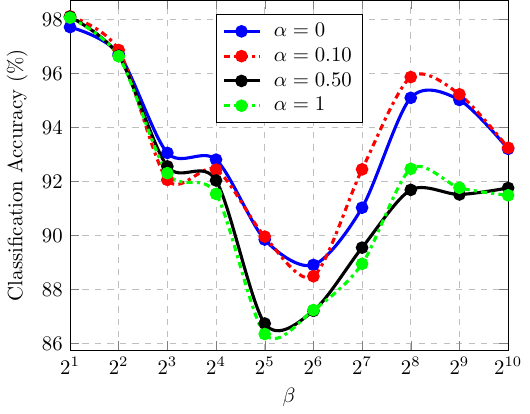}
   \caption{1D + 2D convolutional layers and an LSTM layer trained with SGDM.}
   \end{subfigure}
   \caption{Average classification accuracy. Feature matrices used to train and evaluate the model comprise the weighted logarithmic of $\mathbf{a}_{j,\alpha}$.}
   \label{fig:acc}
\end{figure*}

Figure \ref{fig:acc} shows plots for the average classification accuracy of the proposed method over the hundred subexperiments for two different network architectures: {1D CNN-LSTM and 1D+2D CNN-LSTM}. For each of the two architectures, the CNN-LSTM network is trained separately using ADAM or SGDM. As observed in the plots, the models behave similarly in terms of the scaling factor $\beta$ and the regularization parameter $\alpha$. Generally, the best accuracy is achieved when the coefficient vectors $\mathbf{a}_{j,\alpha}$ are calculated using the Gabor dictionary of atoms with high-time and low-frequency resolution ($\beta = 2^1$) for all values of $\alpha$. Accuracy decreases with increasing value of $\beta$, and the lowest accuracy is recorded at $\beta=2^6$. After that, the accuracy increases with increasing the value of $\beta$.

For most of the resolutions of time-frequency atoms, both CNN-LSTM networks provide the best accuracy when using features calculated with the elastic net regularization for $\alpha=0$ (ridge regression). However, the value for $\alpha=0.1$ competes with ridge regression in providing comparable performance and even better accuracy at some time-frequency resolutions, as is the case of $\beta = 2^1$ and the second architecture with ADAM. These results are consistent with the energy analysis of both the approximation residual $\|\mathbf{x}-\mathbf{D}_j\mathbf{a}_{j,\alpha}\|_2^2$ and the coefficient vectors $\mathbf{a}_{j,\alpha}$ shown in figures \ref{fig:error1} and \ref{fig:energy}. In this context, better classification accuracy is achieved when low-energy coefficient vectors approximate PCG signals with a low-energy approximation residual.

As for the comparison between both architectures, it can be observed that {the model 1D+2D CNN-LSTM trained} using ADAM provides the best accuracy. {In the case of the 1D CNN-LSTM architecture, the results achieved by SGDM and ADM are comparable}. However, classification accuracy decreases when SGDM is used to train {the model 1D+2D CNN-LSTM,} especially for high sparse coefficient vectors $\mathbf{a}_{j,\alpha}$ (large values for $\alpha$) and atoms with moderate time-frequency resolution ($\beta$ equal to $2^6$ or $2^7$). In summary, the best classification accuracy of $98.95\%$ is obtained with {the 1D+2D CNN-LSTM architecture, when using the Gabor dictionary with scale $\beta=2^1$ and regularization parameter $\alpha=0.1$.} 

Tables \ref{tab:res01} and \ref{tab:res31} contain the average values and standard deviations of four classification metrics, namely precision, recall, specificity, and F1 score. These values are measured as a function of selected values for both {$\beta$ and $\alpha$} and for the two model architectures and with the two training optimization algorithms. The average values of the metrics follow a similar trend as that of the accuracy in Figure \ref{fig:acc}. Furthermore, the standard deviation of a given metric decreases as the value of the metric increases. 
\begin{table}
\begin{center}
\caption{Classification metrics and standard deviations (in parenthesis) for the two network architectures when trained with ADAM.}
\begin{tabular}{@{}lllllll@{}}
\hline
\multirow{2}{*}{$\alpha$}&\multirow{2}{*}{\text{Metric}}&\multicolumn{5}{c}{$\beta$}\\
&&$2^1$&$2^2$&$2^4$&$2^7$&$2^{10}$\\
\hline
\multirow{5}{*}{$0$}
&\text{PRE}&$97.90(0.88)$&$97.12(1.00)$&$93.20(1.40)$&$92.06(1.67)$&$94.46(0.96)$\\
&\text{REC}&$97.86(0.90)$&$97.07(1.03)$&$93.07(1.45)$&$91.77(1.75)$&$94.25(0.99)$\\
&\text{SPE}&$99.46(0.22)$&$99.27(0.26)$&$98.27(0.36)$&$97.94(0.44)$&$98.56(0.25)$\\
&\text{F1 score}&$97.85(0.90)$&$97.07(1.03)$&$93.05(1.45)$&$91.76(1.76)$&$94.26(0.98)$\\
\hline
\multirow{5}{*}{$0.10$}
&\text{PRE}&$97.99(0.81)$&$96.92(0.99)$&$90.80(1.77)$&$91.55(1.40)$&$92.98(1.28)$\\
&\text{REC}&$97.96(0.83)$&$96.85(1.01)$&$90.68(1.75)$&$91.28(1.41)$&$92.64(1.36)$\\
&\text{SPE}&$99.49(0.21)$&$99.21(0.25)$&$97.67(0.44)$&$97.82(0.35)$&$98.16(0.34)$\\
&\text{F1 score}&$97.96(0.83)$&$96.85(1.01)$&$90.62(1.78)$&$91.25(1.42)$&$92.65(1.36)$\\
\hline
\multirow{5}{*}{$0.50$}
&\text{PRE}&$97.94(0.90)$&$96.40(0.81)$&$90.63(1.38)$&$90.70(1.58)$&$92.83(1.14)$\\
&\text{REC}&$97.91(0.92)$&$96.34(0.85)$&$90.49(1.45)$&$90.42(1.57)$&$92.40(1.27)$\\
&\text{SPE}&$99.48(0.23)$&$99.09(0.21)$&$97.62(0.36)$&$97.61(0.39)$&$98.10(0.32)$\\
&\text{F1 score}&$97.91(0.92)$&$96.34(0.84)$&$90.43(1.45)$&$90.39(1.59)$&$92.45(1.25)$\\
\hline
\multirow{5}{*}{$1$}
&\text{PRE}&$97.90(0.81)$&$96.73(0.70)$&$90.31(1.39)$&$90.17(1.62)$&$92.58(1.25)$\\
&\text{REC}&$97.85(0.84)$&$96.67(0.73)$&$90.16(1.40)$&$89.89(1.68)$&$92.16(1.39)$\\
&\text{SPE}&$99.46(0.21)$&$99.17(0.18)$&$97.54(0.35)$&$97.47(0.42)$&$98.04(0.35)$\\
&\text{F1 score}&$97.85(0.84)$&$96.66(0.73)$&$90.10(1.41)$&$89.86(1.68)$&$92.21(1.37)$\\
\hline
\multicolumn{7}{c}{The model consists of a 1D convolutional layer and an LSTM layer}\\
\hline
\hline
\multirow{5}{*}{$0$}
&\text{PRE}&$98.05(0.90)$&$97.43(0.94)$&$94.56(1.11)$&$93.76(1.28)$&$94.87(1.27)$\\
&\text{REC}&$98.00(0.93)$&$97.39(0.95)$&$94.47(1.14)$&$93.60(1.30)$&$94.70(1.31)$\\
&\text{SPE}&$99.50(0.23)$&$99.35(0.24)$&$98.62(0.28)$&$98.40(0.33)$&$98.68(0.33)$\\
&\text{F1 score}&$97.99(0.94)$&$97.38(0.96)$&$94.45(1.14)$&$93.59(1.30)$&$94.69(1.33)$\\
\hline
\multirow{5}{*}{$0.10$}
&\text{PRE}&$98.95(0.67)$&$97.19(0.86)$&$92.91(1.44)$&$93.44(1.54)$&$94.18(1.52)$\\
&\text{REC}&$98.95(0.69)$&$97.15(0.87)$&$92.79(1.47)$&$93.26(1.60)$&$93.97(1.58)$\\
&\text{SPE}&$99.74(0.17)$&$99.29(0.22)$&$98.20(0.37)$&$98.32(0.40)$&$98.50(0.40)$\\
&\text{F1 score}&$98.95(0.70)$&$97.14(0.87)$&$92.77(1.49)$&$93.26(1.60)$&$93.97(1.60)$\\
\hline
\multirow{5}{*}{$0.50$}
&\text{PRE}&$98.33(0.65)$&$96.80(1.08)$&$92.42(1.41)$&$93.71(1.47)$&$92.63(1.32)$\\
&\text{REC}&$98.30(0.66)$&$96.74(1.11)$&$92.29(1.42)$&$93.55(1.51)$&$92.37(1.36)$\\
&\text{SPE}&$99.57(0.17)$&$99.19(0.28)$&$98.07(0.36)$&$98.39(0.38)$&$98.09(0.34)$\\
&\text{F1 score}&$98.29(0.66)$&$96.74(1.11)$&$92.26(1.43)$&$93.54(1.51)$&$92.36(1.36)$\\
\hline
\multirow{5}{*}{$1$}
&\text{PRE}&$98.08(0.86)$&$96.91(0.92)$&$92.50(1.47)$&$93.85(1.14)$&$92.72(1.31)$\\
&\text{REC}&$98.04(0.88)$&$96.85(0.93)$&$92.37(1.49)$&$93.72(1.14)$&$92.51(1.32)$\\
&\text{SPE}&$99.51(0.22)$&$99.21(0.23)$&$98.09(0.37)$&$98.43(0.29)$&$98.13(0.33)$\\
&\text{F1 score}&$98.04(0.88)$&$96.86(0.93)$&$92.35(1.51)$&$93.72(1.15)$&$92.51(1.31)$\\
\hline
\multicolumn{7}{c}{The model consists of 1D and 2D convolutional layers and an LSTM layer}\\
\hline
\label{tab:res01}
\end{tabular}
\end{center}
\end{table}
\begin{table}
\begin{center}
\caption{Classification metrics and standard deviations (in parenthesis) for the two network architectures when trained with SGDM.}
\begin{tabular}{@{}lllllll@{}}
\hline
\multirow{2}{*}{$\alpha$}&\multirow{2}{*}{\text{Metric}}&\multicolumn{5}{c}{$\beta$}\\
&&$2^1$&$2^2$&$2^4$&$2^7$&$2^{10}$\\
\hline
\multirow{5}{*}{$0$}
&\text{PRE}&$98.39(0.66)$&$97.59(0.95)$&$94.13(1.12)$&$92.19(1.49)$&$95.59(1.12)$\\
&\text{REC}&$98.37(0.68)$&$97.55(0.96)$&$94.05(1.14)$&$91.99(1.84)$&$95.49(1.16)$\\
&\text{SPE}&$99.59(0.17)$&$99.39(0.24)$&$98.51(0.29)$&$98.00(0.46)$&$98.87(0.29)$\\
&\text{F1 score}&$98.37(0.68)$&$97.55(0.96)$&$94.03(1.15)$&$91.93(1.86)$&$95.49(1.15)$\\
\hline
\multirow{5}{*}{$0.10$}
&\text{PRE}&$97.85(0.79)$&$96.93(0.90)$&$92.04(1.26)$&$91.56(1.25)$&$94.49(1.40)$\\
&\text{REC}&$97.82(0.81)$&$96.88(0.94)$&$91.95(1.25)$&$91.35(1.30)$&$94.34(1.45)$\\
&\text{SPE}&$99.45(0.20)$&$99.22(0.24)$&$97.99(0.31)$&$97.84(0.32)$&$98.59(0.36)$\\
&\text{F1 score}&$97.82(0.81)$&$96.88(0.94)$&$91.94(1.26)$&$91.34(1.29)$&$94.35(1.44)$\\
\hline
\multirow{5}{*}{$0.50$}
&\text{PRE}&$98.04(0.85)$&$97.13(0.65)$&$91.17(1.32)$&$90.32(1.46)$&$93.00(1.78)$\\
&\text{REC}&$98.00(0.88)$&$97.05(0.68)$&$91.08(1.32)$&$90.00(1.50)$&$92.83(1.80)$\\
&\text{SPE}&$99.50(0.22)$&$99.26(0.17)$&$97.77(0.33)$&$97.50(0.38)$&$98.21(0.45)$\\
&\text{F1 score}&$98.00(0.88)$&$97.05(0.68)$&$91.02(1.31)$&$89.98(1.52)$&$92.83(1.80)$\\
\hline
\multirow{5}{*}{$1$}
&\text{PRE}&$97.79(0.80)$&$97.07(0.95)$&$90.70(1.81)$&$89.55(2.02)$&$93.00(1.98)$\\
&\text{REC}&$97.74(0.83)$&$97.00(1.00)$&$90.54(1.83)$&$89.35(2.05)$&$92.77(1.99)$\\
&\text{SPE}&$99.44(0.21)$&$99.25(0.25)$&$97.64(0.46)$&$97.34(0.51)$&$98.19(0.50)$\\
&\text{F1 score}&$97.74(0.82)$&$97.00(1.01)$&$90.49(1.86)$&$89.28(2.08)$&$92.77(1.99)$\\
\hline
\multicolumn{7}{c}{The model consists of a 1D convolutional layer and an LSTM layer}\\
\hline
\hline
\multirow{5}{*}{$0$}
&\text{PRE}&$97.77(0.76)$&$96.78(0.82)$&$92.94(1.26)$&$89.79(1.49)$&$93.47(1.22)$\\
&\text{REC}&$97.72(0.77)$&$96.73(0.83)$&$92.81(1.26)$&$89.03(1.75)$&$93.21(1.32)$\\
&\text{SPE}&$99.46(0.19)$&$99.18(0.21)$&$98.20(0.31)$&$97.26(0.44)$&$98.30(0.33)$\\
&\text{F1 score}&$97.72(0.78)$&$96.72(0.83)$&$92.79(1.27)$&$89.03(1.79)$&$93.21(1.32)$\\
\hline
\multirow{5}{*}{$0.10$}
&\text{PRE}&$98.17(0.99)$&$96.94(1.20)$&$92.58(1.59)$&$92.64(1.54)$&$93.44(1.50)$\\
&\text{REC}&$98.12(1.03)$&$96.87(1.21)$&$92.44(1.62)$&$92.45(1.63)$&$93.25(1.55)$\\
&\text{SPE}&$99.53(0.26)$&$99.22(0.30)$&$98.11(0.41)$&$98.11(0.41)$&$98.31(0.39)$\\
&\text{F1 score}&$98.12(1.03)$&$96.87(1.22)$&$92.42(1.64)$&$92.44(1.64)$&$93.25(1.54)$\\
\hline
\multirow{5}{*}{$0.50$}
&\text{PRE}&$98.13(0.79)$&$96.71(0.78)$&$92.17(1.45)$&$89.60(1.65)$&$92.01(1.63)$\\
&\text{REC}&$98.10(0.81)$&$96.66(0.79)$&$92.04(1.46)$&$89.55(1.75)$&$91.76(1.77)$\\
&\text{SPE}&$99.53(0.22)$&$99.16(0.20)$&$98.01(0.37)$&$97.31(0.44)$&$97.94(0.44)$\\
&\text{F1 score}&$98.10(0.81)$&$96.66(0.79)$&$92.02(1.48)$&$89.24(1.80)$&$91.76(1.78)$\\
\hline
\multirow{5}{*}{$1$}
&\text{PRE}&$98.10(0.75)$&$96.68(0.70)$&$91.69(1.74)$&$88.64(1.60)$&$91.72(1.61)$\\
&\text{REC}&$98.07(0.76)$&$96.64(0.70)$&$91.54(1.75)$&$88.22(1.84)$&$91.49(1.69)$\\
&\text{SPE}&$99.52(0.19)$&$99.16(0.17)$&$97.89(0.44)$&$97.06(0.46)$&$97.87(0.42)$\\
&\text{F1 score}&$98.07(0.76)$&$96.63(0.70)$&$91.52(1.76)$&$88.17(1.85)$&$91.51(1.69)$\\
\hline
\multicolumn{7}{c}{The model consists of 1D and 2D convolutional layers and an LSTM layer}\\
\hline
\end{tabular}
\label{tab:res31}
\end{center}
\end{table}

\begin{figure*}
\centering
\includegraphics[scale=0.55,trim={0cm 0cm 0cm 2cm}]{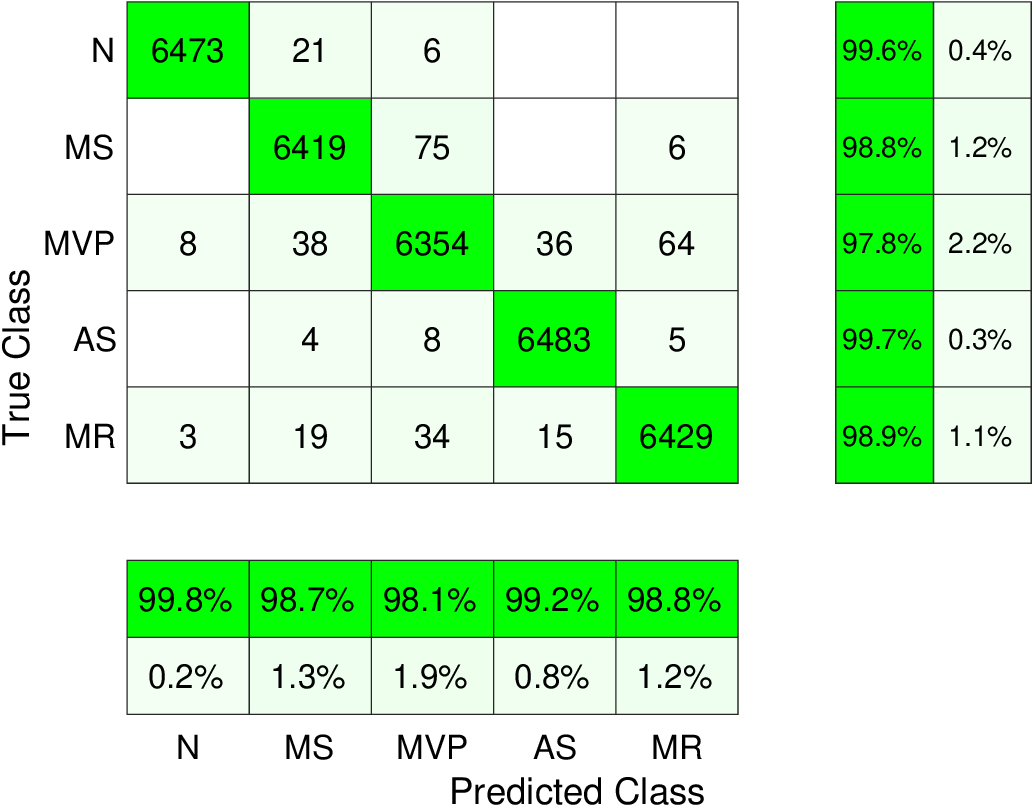}
\caption{The confusion matrix of the five heart valvular conditions for $100$ experiments. It is obtained using the Gabor dictionary constituted with $\beta=2^1$ and using the elastic net regularization with $\alpha=0.1$. The classification model consists of 1D and 2D convolutional layers and an LSTM layer and is trained using ADAM. }
\label{fig:cm}
\end{figure*}
\begin{table*}[h]
\begin{center}
\caption{Classification metrics of the confusion matrix in Figure \ref{fig:cm}.}
\centering
\begin{tabular}{@{}llllll@{}}
\hline
\multirow{2}{*}{\text{Metric}}&\multicolumn{5}{c}{Heart valvular condition}\\
&\text{N}&\text{MS}&\text{MVP}&\text{AS}&\text{MR}\\
\hline
\text{PRE}&$99.83$&$98.74$&$98.10$&$99.22$&$98.85$\\
\text{REC}&$99.58$&$98.75$&$97.75$&$99.74$&$98.91$\\
\text{SPE}&$99.96$&$99.68$&$99.53$&$99.80$&$99.71$\\
\text{F1 score}&$99.71$&$98.75$&$97.93$&$99.48$&$98.88$\\
\hline
\end{tabular}
\label{tab:detmetric}
\end{center}
\end{table*}
Figure \ref{fig:cm} and Table \ref{tab:detmetric} show the confusion matrix and detailed classification metrics, respectively, for the five heart valvular conditions over $100$ experiments {and the best model (1D+2D CNN-LSTM architecture trained with ADAM in combination with Gabor dictionary built with $\beta=2^1$ and $\alpha=0.1$}). It can be observed that PCG signals from healthy hearts (N) are discriminated with the highest performance, followed by valvular disease of aortic stenosis (AS). The classification model provides the lowest metric values in the heart valvular condition of mitral valve prolapse (MVP). Mitral stenosis (MS) and mitral regurgitation (MR) are classified with performance less than that of healthy hearts and AS and higher than that of MVP.

\subsection{Discussion}
The application of deep learning in the development of automated systems for the detection of cardiac abnormalities is the subject of numerous recent studies. The binary classification of PCG as healthy or pathological has been considered highly in many researches. However, there are few studies that have addressed the multiclass classification of PCG signals. Table \ref{tab:table11} compares the proposed method with two recently developed baseline methods using the same dataset analyzed in this study. {The first baseline is based on raw PCG signals or their Fourier components (transformed PCG signals) using a deep CNN-LSTM classification model \cite{AlIssa2022ALH} that consists of a complex network composed of $4$ convolutional layers, $4$ batch normalization layers, $4$ ReLU units, $3$ max-pooling layers, and $2$ LSTM layers and is trained using ADAM with $10$-fold cross-validation.} 

The second baseline method is based on variational mode decomposition for feature extraction and a light CNN-LSTM model for feature classification \cite{mfakhryvmd}. PCG signals are decomposed into a finite set of intrinsic mode functions using VMD, and classification features are obtained by applying a weighted logarithmic operation to the mode functions. The CNN-LSTM light model comprises one convolutional layer, one ReLU unit, and one LSTM layer and is trained using SGDM $1000$ times separately, using random selection for training and testing recordings. 

The table also shows the accuracy of the proposed network architectures, namely 1D CNN-LSTM and 1D+2D CNN-LSTM using the two training optimization algorithms, namely SGDM and ADAM, in the optimal combinations of the resolution parameter ($\beta$) and the regularization parameter ($\alpha$). It is observed that the networks perform well when Gabor dictionaries of atoms with high-time resolution are used to represent the PCG signals. Moreover, the classification accuracy improves by introducing some sparsity in the coefficient vectors $\mathbf{a}_{j,\alpha}$, which is confirmed by the high accuracy obtained when the elastic net regularization with $\alpha=0.1$ is used to approximate the signals. Moreover, large networks provide better performance when trained with ADAM, while SGDM is suitable for training small networks. 

{As for the comparison with the other state-of-the-art systems, on the one hand, it can be observed that our best model achieves a relative classification error reduction of $30.92~\%$ with respect to the first baseline with raw PCG signals. This result suggests that features with high discrimination properties need neural networks with less complexity and fewer parameters to train than raw signals or less discriminative features. On the other hand, the relative error reduction with respect to the second baseline is of $22.22~\%$. This outcome indicates that our proposed features derived from Gabor dictionaries and elastic net are more suitable to model PCG signals than the VMD technique.}

\begin{table}
\centering
\caption{Comparison of the proposed method and baseline methods.}
\label{tab:table11}
\begin{tabular}{@{}lllll@{}}
\hline
Ref. & Features	& Classifier & Training & Accu.\\
\hline
\multirow{2}{*}{\cite{AlIssa2022ALH}} & Raw PCG signals & \multirow{2}{*}{Deep 2D CNN-LSTM} & \multirow{2}{*}{ADAM} & $98.48$\\
& Transformed PCG signals&&& $95.40$\\
\cite{mfakhryvmd}  & VMD + weighted logarithmic & Light 1D CNN-LSTM& SGDM & $98.65$ \\
\multirow{2}{*}{Prop.} & Gabor dictionary ($\beta=2^1$)+ & \multirow{2}{*}{1D CNN-LSTM} & \multirow{2}{*}{ADAM} & \multirow{2}{*}{$97.97$}\\ 
&elastic net ($\alpha=0.1$)&&\\
\multirow{2}{*}{Prop.} & Gabor dictionary ($\beta=2^1$)+ & \multirow{2}{*}{1D CNN-LSTM} & \multirow{2}{*}{SGDM} & \multirow{2}{*}{$98.39$}\\ 
&elastic net ($\alpha=0$)&&\\
\multirow{2}{*}{Prop.} & Gabor dictionary($\beta=2^1$)+ & \multirow{2}{*}{1D+2D CNN-LSTM} & \multirow{2}{*}{SGDM}& \multirow{2}{*}{$98.14$}\\ 
&elastic net ($\alpha=0.1$)&&\\
\multirow{2}{*}{Prop.} & Gabor dictionary($\beta=2^1$)+ & \multirow{2}{*}{1D+2D CNN-LSTM} & \multirow{2}{*}{ADAM}& \multirow{2}{*}{$98.95$}\\ 
&elastic net ($\alpha=0.1$)&&\\
\hline
\end{tabular}
\end{table}
\section{Conclusion and future work}
This article presents an approach to better model heart sound signals by optimizing both the resolution of time-frequency atoms and the regularization of fitting models. To achieve this goal, the fitting models have been obtained by applying elastic net regularization of linear models with an overcomplete dictionary of Gabor atoms to heart sound signals. In addition to that, we have proposed to exploit the fitting models to develop time-frequency feature matrices with a sparse representation. This approach allows us to better capture the time-frequency structure of heart sound signals while preserving the underlying sparsity of the data. 

We have assessed the performance of the proposed features over a publicly available database of heart sounds for the task of cardiovascular disease classification. For this purpose, we have designed two different deep neural network architectures. The first system consists of a 1D convolutional layer and a long short-term memory (LSTM) layer, while the second one is composed of 1D and 2D convolutional layers and an LSTM layer. Both networks have been trained with two different algorithms, namely stochastic gradient descent with momentum (SGDM) and adaptive moment (ADAM), using the aforementioned time-frequency matrices as input features.

We have conducted extensive experiments over these two network architectures with input feature matrices obtained with different resolution and regularization parameters. The best classification accuracy of $98.95\%$ is obtained with the second network architecture when high-time and low-frequency resolution matrices are used as input features. This second architecture, which consists of 1D and 2D convolutional layers and an LSTM layer, is more complex but provides better performance. This finding suggests that deeper and more complex neural network architectures can better capture the complex structures in the time-frequency matrices and improve classification accuracy.

{In future work, we are planning to extend the proposed methodology by substituting the real Gabor dictionary with a complex Gabor dictionary, as we think that using a dictionary of complex atoms will result in sparser approximations of PCG signals, which will highly benefit the classification performance. In addition, we will explore and evaluate the performance of pre-trained deep models using transfer learning in conjunction with STFT or with the proposed new class of time-frequency representations and compare them to the designed CNN-LSTM network.}  

\section*{Acknowledgement}
The authors acknowledge the support of the Spanish State Research Agency (MCIN/AEI/10.13039/5011000110) through project PID2020-115363RB-I00.

\bibliographystyle{elsarticle-num} 
\bibliography{sn-bibliography}

\end{document}